# Gas puff imaging of plasma turbulence in the magnetic island scrape-off layer of W7-X

S.G. Baek[1], S. Ballinger[1], O. Grulke[2,3], C. Killer[2], A. von Stechow[2], J. L. Terry[1] F. Scharmer[2,4], and B. Shanahan[2]

[1]MIT Plasma Science and Fusion Center, Cambridge, USA
[2]Max Planck-Institut für Plasmaphysik, Greifswald, Germany
[3]Technical University of Denmark, Lyngby, Denmark
[4]University of Greifswald, Greifswald, Germany

Abstract: The turbulence characteristics of the scrape-off-layer (SOL) plasma in the W7-X stellarator are investigated using a gas-puff-imaging (GPI) diagnostic, newly installed and operated during the OP 2.1 campaign. The SOL plasma on W7-X features a set of island divertors for heat and particle exhaust and provides a unique environment for studying SOL turbulence and transport. This paper focuses on the O-point region of the magnetic island divertor SOL in the standard magnetic configuration. Fourier and cross-correlation analyses show that turbulence flows are predominantly in the poloidal direction (i.e., direction tangent to the last closed flux surface) with significantly weaker radial motion. This suggests dominant ExB convection and suppressed filamentary transport compared to those observed in the far scrape-off-layer region of tokamaks, as further supported by high-resolution skewness and kurtosis data that show the absence of intermittent, bursty filamentary events. Additionally, a relationship between the radial profile of the connection length and the sheared poloidal flow structure is reported, suggesting a possible linkage among magnetic topology, turbulence dynamics, and turbulence generation.

1. Introduction

W7-X is an optimized superconducting stellarator ($B_0$ = 2.5 T, $R_0$ = 5.5 m, a = 0.53 m), which investigates the physics and engineering of the stellarator reactor concept [1]. It utilizes the magnetic island divertor concept. In the "standard" magnetic configuration that is commonly used on W7-X, five magnetic islands surround the core plasma. Ten modular divertor targets intercept the magnetic islands and provide transport channels for particle and power exhaust. Transport processes within the island plasmas are an active research area.

In the W7-X scrape-off-layer (SOL), the long connection length $L_c$ and small pitch on the order of $B_r/B \approx 10^{-3}$ [2] make perpendicular transport important. One such channel is intermittent convective transport due to field-aligned filamentary structures [3–6]. They are often termed blobs since that is their appearance when imaged onto the 2D poloidal-radial plane. On tokamaks, they are measured to be created near the last closed flux surface (LCFS) and exhibit a radial ballistic motion. Initial observations and simulations on W7-X [7,8] indicate that this transport channel may not be dominant due to a weak curvature drive with a large major radius, as evidenced by a slow radial motion (~ 100 m/s) and a short lifetime. However, their generation, propagation, fate within the 3D island geometry, and their contribution to transport are not clearly established.

Another cross-field transport channel is the convection of turbulence by the drift flows (chiefly poloidal), which is gaining attention on W7-X. For example, poloidal ExB motion is believed to be an important element in SOL transport [9,10]. Furthermore, several oppositely directed poloidal flows within the magnetic island are reported, and the characterizations of the dependence of the flow magnitude on various plasma parameters are progressing. Experimental characterization of turbulence transport within the magnetic island region of the W7-X SOL plasma contributes to understanding of the interplay between the 3D island divertor geometry and the cross-field turbulence properties, ultimately advancing the fundamental understanding of the island divertor operation.

In this paper, a gas puff imaging (GPI) diagnostic [11] is used to report turbulence properties and dynamics in the magnetic island region of the W7-X SOL plasma, focusing on the O-point region of the magnetic island in a W7-X standard magnetic configuration. The closed-field-line regions in the SOL provide a unique environment for studying turbulence and transport. GPI is a well-established diagnostic for studying edge and SOL turbulence fluctuations on tokamaks [12], and has recently been deployed on W7-X to diagnose and understand the 3D island diverter SOL. As a part of the initial analysis of the GPI experiment, the following three topics are presented. First, the radial profile of poloidal flow is evaluated based on Fourier analysis. The observed shear layers are qualitatively interpreted in terms of the connection length, $L_c$, profile. Second, 2D cross-correlation analyses are presented to determine the poloidal and radial motions of turbulence structures, indicating the radial motion, which is associated with filamentary transport, is significantly slower than poloidal motion. Finally, to further characterize filamentary fluctuations on W7-X, turbulence statistical properties are examined, showing a negligible level of skewness and excess kurtosis than those observed in the tokamak far-SOLs.

2. Gas Puff Imaging on W7-X and Experimental Condition

GPI measures the fluctuations in line emission due to the plasma interaction of the neutrals that are locally puffed into the target plasma edge region. The emission intensity is a function of the neutral density $n_0$, the electron density $n_e$, and the emission rate coefficient function, $f(n_e, T_e)$, which is a function of the local electron density and temperature. The emission fluctuation represents a combination of electron density and temperature fluctuations. The three major GPI hardware components on W7-X include (i) a supersonic nozzle system that puffs the $H_2$ neutral gas into the target field of view (FoV) region and (ii) an imaging and relaying optics system that collects and transmits the $H_\alpha$ line emission (656 nm) to (iii) an APDCAM-10G camera with an 8x16 array of avalanche photodiode detectors. It records the images at 2 MSamples/sec. Please refer to [11] for a more detailed description of the GPI system on W7-X.

Figure 1 shows the time history of the reference discharge, XP20230323.039, studied in this paper. The line integrated density is constant at $n_{el} = 8 \times 10^{19}$ m$^{-2}$ (or the line-averaged density of $6 \times 10^{19}$ m$^{-3}$), and electron cyclotron (EC) heating is stepped down from $P_{EC} = $ 4 MW to 2 MW in a hydrogen plasma. Two hydrogen GPI puffs are made at t = 2.1 and 5.1 seconds at two EC power levels, and the data taken during these two periods are analyzed. Each GPI gas puff amount (~3.3 mbar-l) is small relative to the main plasma fueling (a total of ~200 mbar-l for the duration of the discharge) and the density is feedback-controlled, so the puffs have little effect on the line-integrated density. The GPI puff valve open duration is 50 msec. Note that a core fueling rate is at a level of ~30 mbar-l/sec, and the GPI puff flow rate for the given plenum backing pressure of 1000 mbar is ~8 mbar-liter/sec during the steady-state phase of the GPI gas puff. During the second puff, the radiated power is ~80% of the injected power, approaching toward a divertor detached state.

Figure 2 (a) shows the poloidal cut of the W7-X plasma at the toroidal location of the GPI nozzle with the GPI field-of-view (FoV) marked in red. The five independent island chain is visible in the figure that also shows the GPI view location that is centered on the island located at the outboard midplane. Figure 2 (b) shows the zoomed-in view of the GPI FoV region, as defined with the individual 8×16 detector pixel center positions marked in red. The distance between the LCFS and the innermost edge of the FoV is 10 mm, based on the pixel at the top row. Two GPI nozzles are approximately 10 cm major-radially further behind from the O-point of the island. The O-point of the island is located at the 5th column of the view from the innermost side and the 4th row from the top in the FoV. Figure 2 (c) shows the 2D color-coded connection length $L_c$ profile. Here, the connection length is the sum of the two distances that a magnetic field line travels from the GPI measurement plane ($\varphi = 274°$ in the W7-X coordinate system) toroidally clockwise and counterclockwise until it intersects the divertor surface. The closed field line regions are denoted in white colors. They are seen inside the LCFS, and also around the O-point in the center of the island, which is not intersected by a divertor target. The centers of each GPI 8x16 pixel view are overlaid as well. Each column of the pixel views is nearly tangent to the LCFS, providing 16 "poloidal" views across each of the 8 different "radial" columns. In this study, the radial direction is defined as the normal to the LCFS along the row direction, with the understanding that the closed-field-line curvatures, particularly near the O-point, does not result in perfect alignment between the flux surface and the column views. Figure 3 shows the 2D color-coded connection length $L_c$ in the forward (reverse) magnetic field direction. On W7-X, the default B-field direction is counterclockwise when viewed from the top. The corresponding maps for the diverter target modules intercepted by the field lines traced from the GPI measurement plane are also provided.

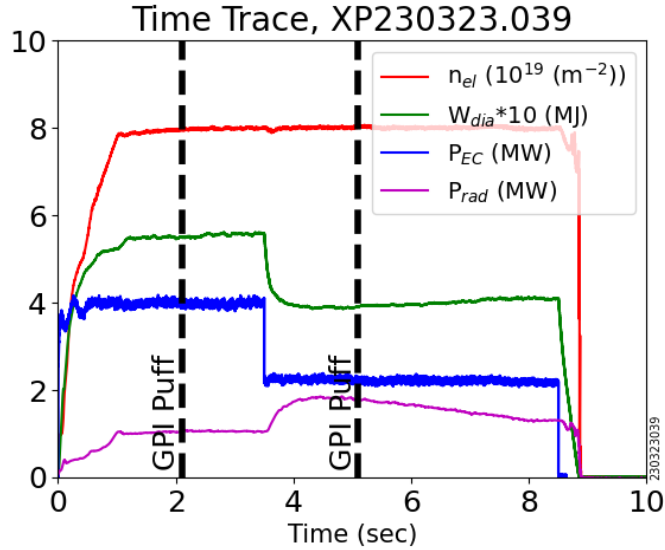

Figure 1. Time trace of key plasma parameters for discharge XP230323.039, including the line-integrated density (red), plasma stored energy, $W_{dia}$ (green), electron cyclotron power, $P_{EC}$ (blue), and total radiated power, $P_{rad}$ (purple). GPI puff events are marked by the black vertical dashed lines at t = 2.1 and 5.1 seconds.

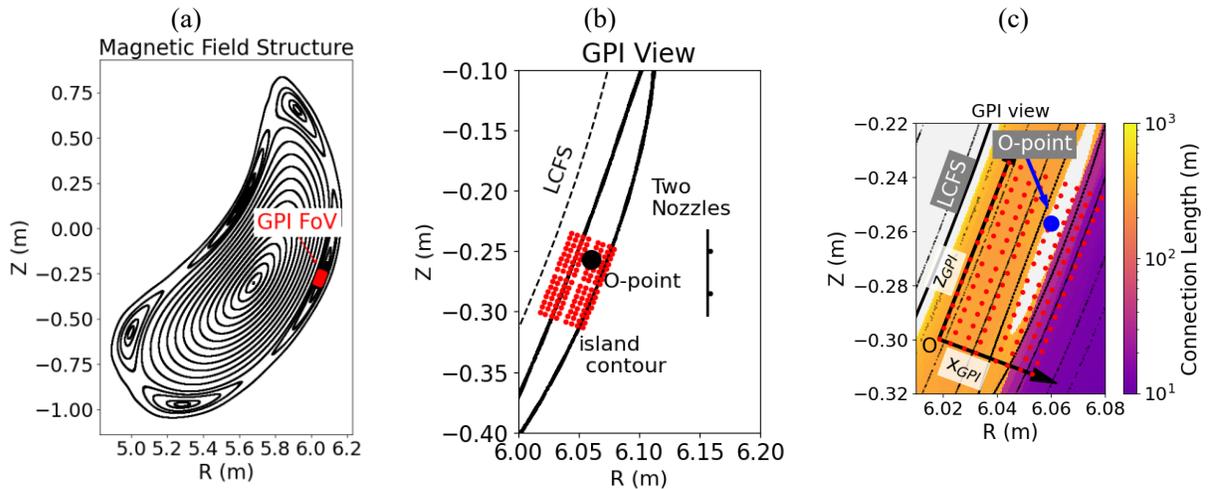

Figure 2. (a) Poincaré plot that shows the magnetic field geometry at the toroidal location of the GPI nozzle (φ = 274° in the W7-X coordinate system). The GPI field of view (FoV) region is indicated by the red rectangle. (b) The zoomed-

in view around the GPI FoV, LCFS, O-point of the magnetic island, and an exemplary island flux surface contour. Each red dot represents a center view of the detector's 8x16 pixel array. (c) 2D color-coded total connection length $L_c$ profile, with the GPI pixel views overlaid. The GPI local coordinate system ($x_{GPI}$ and $z_{GPI}$) is also shown.

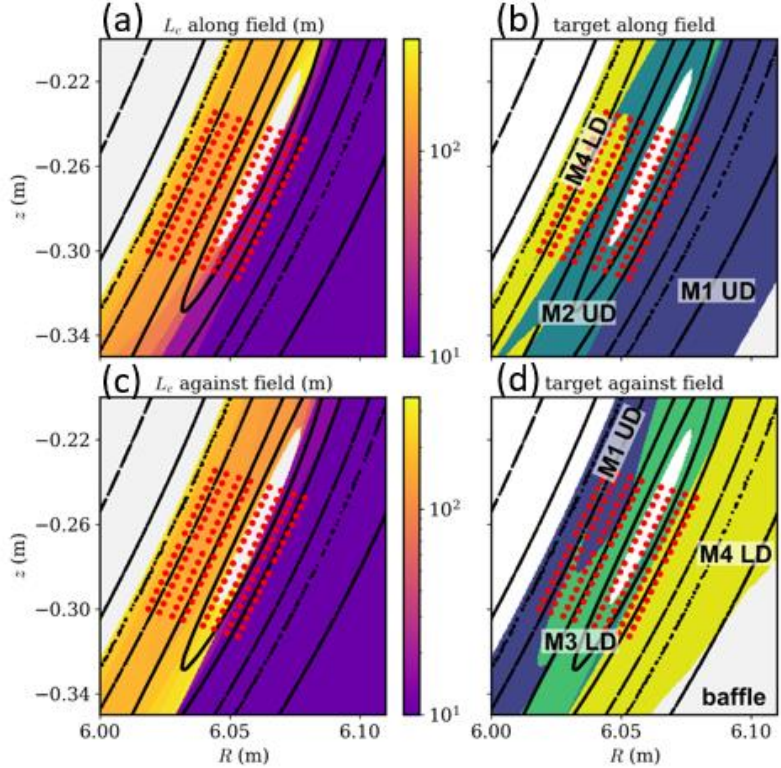

Figure 3. (a,c) 2D color-coded connection length $L_c$ in the forward (reverse) magnetic field direction, with the GPI pixel views overlaid. On W7-X, the default B-field direction is in the counter clockwise direction when viewed W7-X from the top. (b,d) Diverter target modules intercepted by the field lines traced from the GPI measurement plane ($\varphi = 274°$) in the forward (reverse) magnetic field direction. Here, M1 UD refers to Module 1 Upper Divertor, M4 LD refers to Module 4 Lower Divertor, and so forth.

3. Correlation between connection length transition and poloidal velocity shear

In this section, the relationship between poloidal flow dynamics and magnetic topology in the W7-X SOL is reported by correlating the connection length transition to the poloidal velocity shear layers. Figure 4 (a) shows the 1D cut of the connection length profile along the "radial" (or, row) direction. The x-axis represents the distance normal to the LCFS. In the figure, the closed field line region of the magnetic island can be seen with a radial width of ~10 mm, with the O-point situated near column 5 of the GPI FoV ($x_{GPI} - x_{LCFS} = 32.7$ mm). An abrupt reduction in the connection length from infinity to less than 30 m is also observed in the radially outboard side of the magnetic island, which corresponds to the target shadow region. Note that the target shadow region is a plasma region shielded from parallel transport by the discontinuous target plates [13]. While this region is inaccessible to the island SOL plasma through parallel transport, cross-field transport allows heat and particles to enter this target shadow region, potentially reaching plasma-facing components not designed to absorb power. The steep change in connection length here highlights the large changes in magnetic topology within the GPI FoV. For this plasma, there are no density and temperature profile measurements available across the O-point of the island. For a typical density and temperature profile in the island region, please refer to the profile measurements by a scanning probe system in [12, 13] or a thermal helium beam diagnostic [14]. A typical range of the density and temperature values in the magnetic island SOL is in ~ 5 - 20 x$10^{18}$ m$^{-3}$ and 5 ~ 30 eV. As a side note, while the discharge being analyzed here is in the high density range ($n_{el}$ = 8x$10^{19}$ m$^{-2}$), a recent report [14,15] indicates a hollow temperature profile with a minimum at the island O-point, at low densities ($n_{el}$ < 4x$10^{19}$ m$^{-2}$).

Figures 4 (b) and (c) show the radial profile of the poloidal velocity distribution at $P_{EC}$ = 4 MW and 2 MW, respectively. To produce the radial profile of the histogram-like distribution of the poloidal velocity based on Fourier analysis [15], the procedure is as follows. After a 2D temporal and poloidal spatial transform, the fluctuation Fourier power spectral density, s($k_p$,f), is normalized at each frequency, yielding a 2D frequency-normalized spectrum: s($k_p$| f) = s($k_p$,f) / $\sum_i s(k_{p,i}, f)$. This normalization is to highlight and help identify fluctuation features by compensating for the rapid fall-off of the power spectral densities at high frequencies. Then, the distribution of the poloidal phase velocity, $v_{ph}$ = 2πf/$k_p$, is obtained by integrating the s($k_p$| f) values along each line of constant phase velocity, resulting in the distribution of the poloidal phase velocities: d($v_{ph}$) = $\sum_i s(k_{p,i}|f_i)$ where the sum is over the components that meet the condition of $v_{ph}$ = 2π$f_i$/$k_{p,i}$ at each radial location. The exact value of the magnitude, d($v_{ph}$), named here as the pseudo fluctuation spectral power density, is not a quantity of interest because its magnitude does not strictly represent the actual fluctuation power spectral density. Rather, it is the poloidal velocity itself, and this technique helps visualize and identify the radial profile of the dominant poloidal phase velocity and its spread in poloidal velocity space. This approach is also useful in identifying the presence of shear layers because counter-propagating features can be easily identified.

As shown in Figure 4 (b), the poloidal velocity histogram exhibits distinctive features at the velocities of approximately ±2 km/s, with a finite spread. In the W7-X SOL, the poloidal velocity is expected to be driven by the $E_r$xB drift flow, as evidenced by the directional flip with a reversal of the toroidal magnetic field. In the figure, three shear layers are observed across the GPI radial coverage: between columns 1 and 2, between columns 4 and 5, and between columns 6 and 7. Such profiles of the sheared poloidal flow appear to correlate with the transitions in the $L_c$ profile. For example, the second shear layer between columns 4 and 5 aligns with the location near the steep increase of the connection length from ~300 m in the main SOL (i.e., between the LCFS and the confined island) to infinity in the confined island. Similarly, the shear layer between columns 6 and 7 is near the radial location where $L_c$ sharply decreases.

Figure 4 (c) shows the poloidal velocity distribution at a reduced input power of 2 MW. The magnitudes of the dominant phase velocities are reduced to about ±1 km/s, and its spread becomes narrower than the high power case. With a reduced input power, the temperature reduction in the SOL is expected, which is likely to weaken the temperature-gradient induced radial electric field $E_r$xB drive. For example, no noticeable features are seen in the target shadow region in columns 7 and 8. Nevertheless, the general pattern of shear layer locations with respect to the connection length transition location remains unchanged, particularly for the shear layer between columns 4 and 5, despite the reduction in input power.

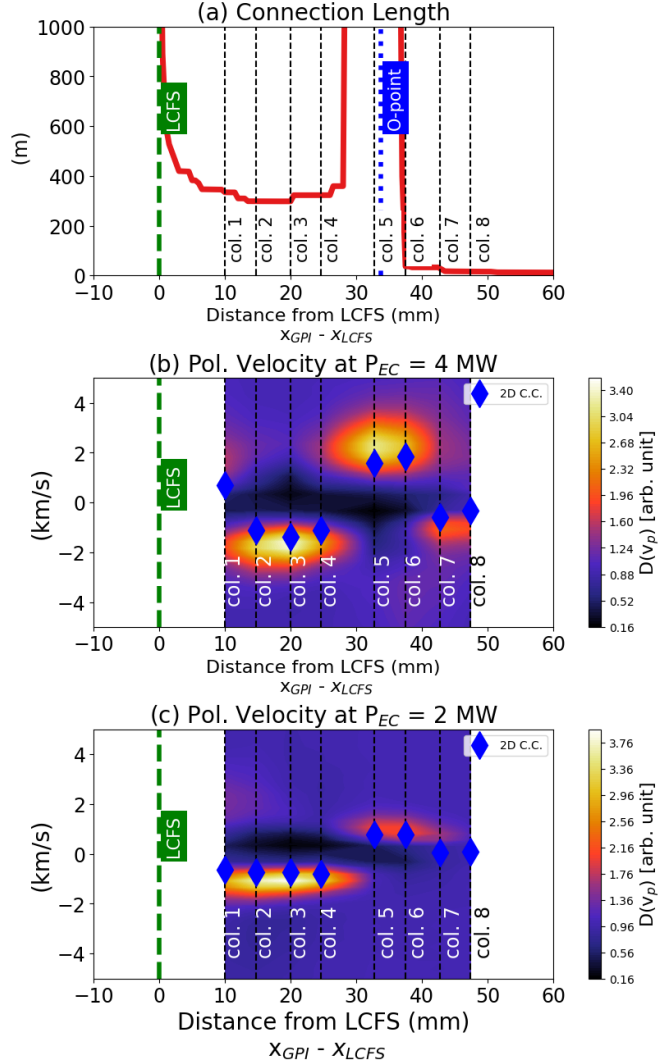

**Figure 4.** (a) The connection length, $L_c$, profile across the radial coordinate. The radial coordinate is normalized to the distance normal to the LCFS. While the maximum connection length shown in the figure is limited at 1 km, the closed field regions effectively have infinite connection lengths. The 2D radial-, velocity-space contour plots of the distribution of poloidal velocity as obtained from Fourier analysis are shown for the (b) $P_{EC}$ = 4 MW and (c) 2 MW cases. The blue diamonds in (b) and (c) are the poloidal velocities evaluated from 2D cross-correlation, discussed in Section 4.

Such an alignment between the shear layer location and the connection length profile can be qualitatively understood in terms of the 2D potential structure, particularly for the second and third shear layers (between columns 4 and 5 and between columns 6 and 7). Recent reciprocating probe measurements [16], which reconstruct the radial electric field from floating potential measurements, show a positive potential around the closed field-line region of the magnetic island, sandwiched by a negative potential structure. Interestingly, such a 2D potential structure follows the shape of the magnetic island geometry. While there is a qualitative agreement of the radial profile of the poloidal velocities between the probe and GPI measurements, we also find the radial offset of ~ 1 cm regarding the shear layer locations between the two diagnostics. Further details will be reported in [15]. The underlying mechanism that determines the 2D potential profile and its relation to the $L_c$ profile needs further investigation with a direct comparison with theory and modeling [7].

4. Flow analysis using 2D cross-correlation

In this section, 2D cross-correlation analyses are applied to compare the poloidal velocity to that from the Fourier analyses and to attempt to determine radial motions. The 2D cross-correlation method can be advantageous because standard two-point time delay estimates are often prone to the barber pole effect, especially when only a limited number of pixels (e.g, two pixels) are involved [17–19]. For example, this effect can result in the apparent registration of a radial velocity (left-right) in the 2D radial-poloidal plane, even when an elongated and tilted feature is moving solely in the poloidal direction (up-down). In the 2D cross-correlation [20,21], the time-series signal is cross-correlated against another time-series signal to evaluate the similarities between the two. For the two signals, $A(x_1,y_1,t)$ and $A(x_2, y_2, t + \tau)$ with a time lag $\tau$, the normalized cross-correlation function is defined as follows:

$$C(x_1, y_1, x_2, y_2, \tau) = \frac{\sum_i [A(x_1, y_1, t_i) - \bar{A}(x_1, y_1)][A(x_2, y_2, t_i + \tau) - \bar{A}(x_2, y_2)]}{\sqrt{\sum_i [A(x_1, y_1, t_i) - \bar{A}(x_1, y_1)]^2 \sum_i [A(x_2, y_2, t_i) - \bar{A}(x_2, y_2)]^2}}$$

The bar denotes a time average. The brightness data is normalized by subtracting the mean brightness to eliminate the pixel-to-pixel variation. The sum is over the time series data over the window. Further, the cross-correlation function is normalized by the square root of the variances. The cross-correlation function is a dimensionless quantity, where $C = +1$ denotes a perfect positive correlation, and $C = 0$ indicates that two signals are uncorrelated. As shown by the blue diamonds in Figure 4 (b) and (c), the poloidal velocities obtained by this technique match well with those obtained by the Fourier method.

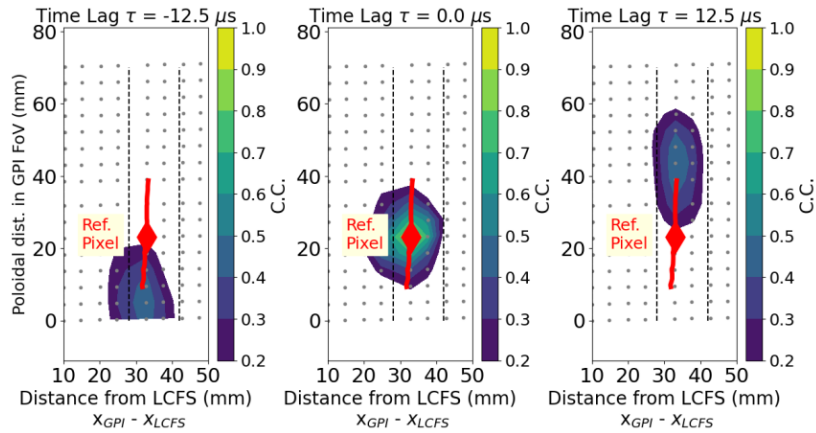

**Figure 5.** Color-coded 2D cross-correlation function values for the discharge XP230323.039 at t = 2.1 – 2.15 sec for three different lag times for the reference pixel is at [column, row] = [5,11], marked as a red diamond. This reference column location spans across the O-point region ($x_{GPI} - x_{LCFS} = 32.7$ mm). The black dashed lines denote the shear layer locations, as shown in Figure 4. The red curve is the trajectory of the peak of the cross-correlation function across the time lag from -12.5 μs to +12.5 μs.

Figure 5 shows the 2D cross-correlation analysis applied between the reference pixel at (column, row) = (5,11), marked as a red diamond in the figure, and the remaining pixels. This reference column spans across the O-point ($x_{GPI} - x_{LCFS} = 32.7$ mm). The time-series brightness signals are bandpass filtered (3 - 800 kHz) before applying the cross-correlation. The 50 msec-long data are segmented into 5 windows, and averaged. The resulting 2D cross-correlation function is interpolated in the GPI FoV plane (~ 41 mm x ~75 mm) by subdividing it into a 200 x 200 grid. Note the spatial extent of the cross-correlation function in each time lag remains approximately 1 cm in diameter seen at zero time delay. The averaged trajectory of the peak of the cross-correlation function is indicated by a red curve. The upward motion of the turbulence structure is apparent, consistent with the Fourier analysis. No clear radial motion is registered. By tracking the radial displacement of turbulence features, the radial location of the maximum cross-correlation function hints at a steady but much slower motion at ~80 m/s ($\Delta x_{GPI} \approx 2$ mm over 25 μs).

To check the consistency of this small level, the cross-correlation analysis is repeated for other reference rows at the same fixed reference column 5, as shown in Figure 5 (a). Figure 5 (b) shows the poloidal displacement of the maximum of the cross-correlation function across the different reference rows. In all cases, the steady upward slopes are seen, clearly indicating the poloidal propagation ($v_{pol} \approx 1.5$ km/s). Figure 5 (c) shows the radial displacement, showing minimal radial motion that is, at most, $v_{rad} \approx 85$ m/s, which is significantly slower than the poloidal motion. It appears that a finite radial propagation is clearer for the reference rows further below from the O-point. We note that there is some uncertainty in the actual value of the radial motion since it is being evaluated in the presence of poloidal motion that is roughly 15 times greater and therefore requires that the assignment of the R and Z coordinates to fluctuation feature, as it propagates through the GPI FoV, be exact. While it is shown [10] that the radial smearing of the brightness measurements, arising from the finite gas cloud size and chordal integration, is expected NOT to cause a systematic change in the mapping of the R coordinate of the actual turbulence feature to the R coordinate assigned in the images, the diagnostic instrumental resolution is limited to ~5.5 mm poloidally and ~10 mm radially. Therefore, we cannot be sure of this at the sub-mm resolution applied in the radial tracking that determines $v_{rad}$. If a structure moves one pixel (~0.005 m) over 25 μsec, this would be equivalent to 200 m/s, suggesting that the radial motion is likely less than 200 m/s in this island region.

Figure 7 shows the resulting quiver plot of the velocity based on 2D cross-correlation. The "poloidal" and "radial" velocities (or the components tangent and normal to the LCFS) are linearly evaluated by taking the 2D positions of the maximal CCF at the time lags between $\tau = -7.5$ μs and $\tau = +7.5$ μs. The peak of the cross-correlation function remains above 0.5 in the 15 μs time window. Results from every other two rows are shown. The result shows the very dominant poloidal motion with the shear layer features observed in the Fourier analysis. The radial component is evaluated to be slightly higher in the target shadow region, but should be taken with caution due to a narrower time window of higher correlation (the velocity estimates in the last column are not shown). In general, the 2D cross-correlation confirms a very slow radial motion of turbulence within the magnetic island region of the W7-X SOL at least when compared to the poloidal motion there.

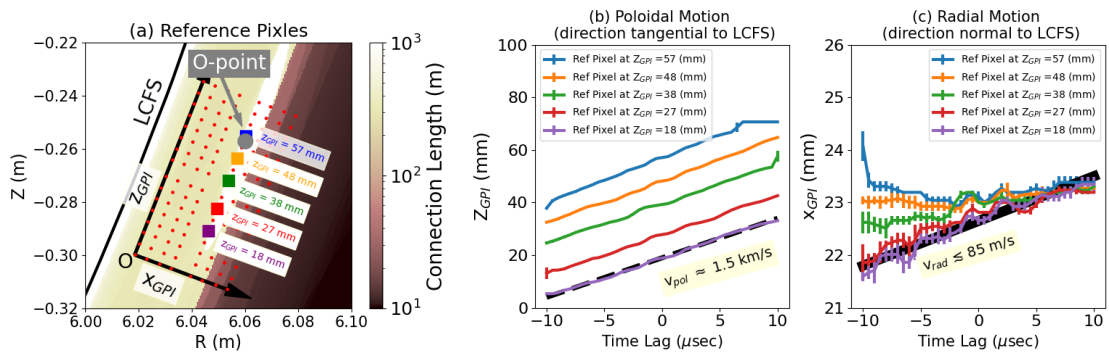

**Figure 6. Tracking of the radial and poloidal locations of the maximum correlation function values evaluated at different time lags using the five different reference rows at the fixed column number 5 ($x_{GPI} - x_{LCFS} = 32.7$ mm). (a)**

Five reference pixels are shown for visual guidance, along with the GPI coordinate system ($x_{GPI}$ and $Z_{GPI}$), which is normalized to the position of the lowermost, leftmost pixel. The reference column spans the O-point region. The centers of each of GPI's 8x16 pixel views are indicated by the red circles. (b) Poloidal displacements as a function of time lag, showing a consistent poloidal velocity at $v_{pol} \approx$ 1.5 km/s. (c) Radial displacements, indicating slower radial motion $v_{rad} \approx$ 85 m/s at most.

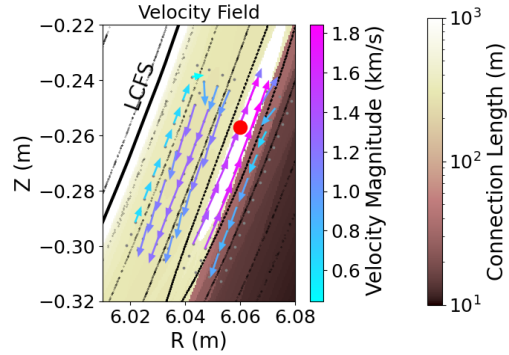

**Figure 7.** Evaluation of 2D velocities of turbulence feature based on the 2D cross-correlation analysis shown in Figures 5 and 6. The velocities are represented as a quiver plot, with vectors shown for every other row. Overlaid is the connection length map, ranging from 10 m to 1000 m. The O-point of the island is marked by the red circle, and the LCFS contour is indicated by the thick black curve. The centers of each of GPI's 8x16 pixel views are indicated by the gray circles.

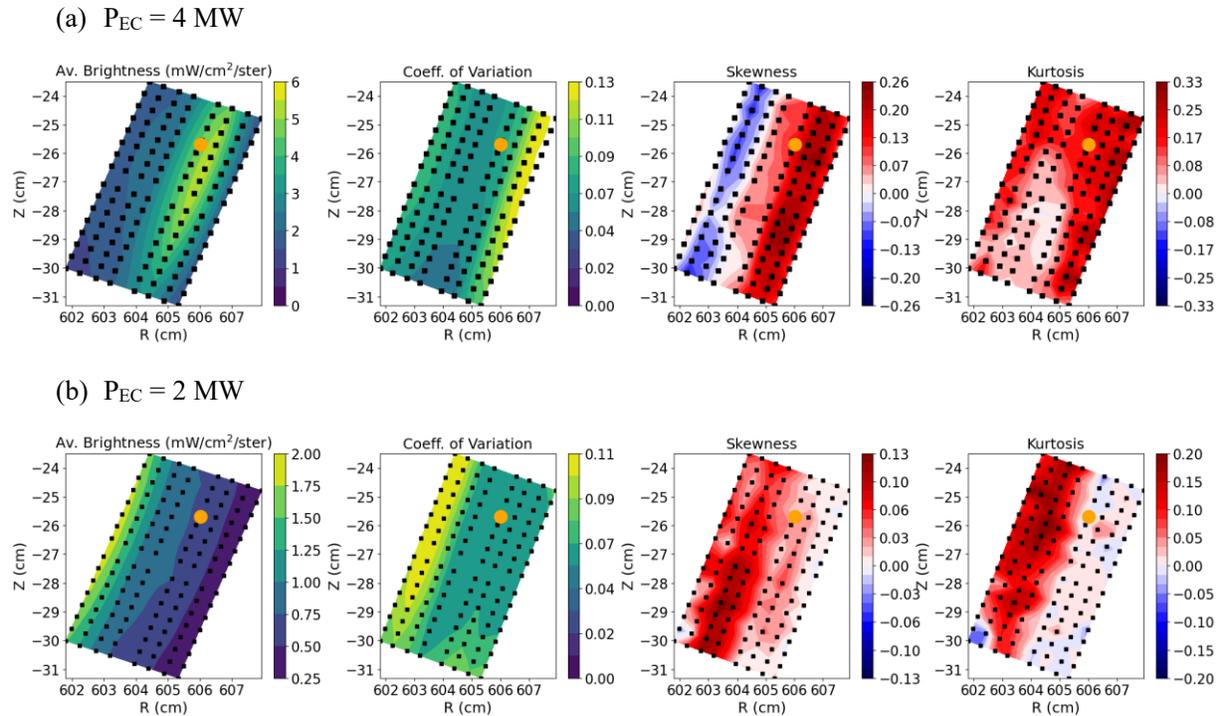

(a) $P_{EC}$ = 4 MW

(b) $P_{EC}$ = 2 MW

**Figure 8.** 2D profiles of the average brightness, coefficient of variation (standard deviation over mean), skewness, and kurtosis at (a) $P_{EC}$ = 4 MW and (b) $P_{EC}$ = 2 MW. The centers of each of GPI's 8x16 pixel views are indicated by the black rectangles. The O-point location of the magnetic island is indicated by an orange circle.

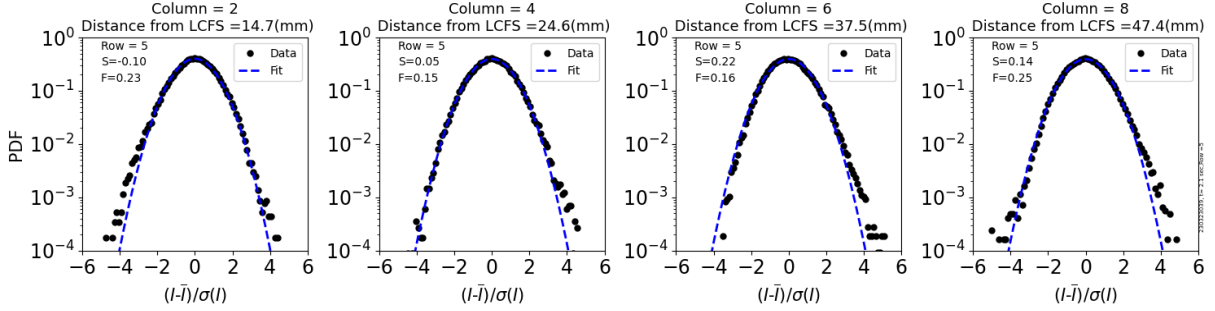

**Figure 9.** Four probability Density Functions (PDFs) of the normalized brightness signals in the $P_{EC}$ = 4 MW case across the four columns for the pixel views at $x_{GPI}$ - $x_{LCFS}$ = 14.7, 24.6, 37.5, and 47.4 mm in the GPI FoV plane at $Z_{GPI}$ = 52 mm (5th row from the top). A best-fit curve to the Gaussian distribution with the forced mean value of 0 is shown in a blue dashed curve in each panel. In the label, S (F) is the symbol for skewness (kurtosis).

5. Correlation between poloidal velocity shear and turbulence statistics

To further investigate this small radial transport, this section focuses on the statistical characterization of fluctuations, specifically skewness and kurtosis. These higher-order moments provide insights into the asymmetry of the fluctuation dynamics and intermittency of fluctuations. These have been studied and characterized extensively in tokamaks and are associated with radial filamentary transport in the tokamak far SOL. Figure 7 shows the 2D profiles of mean brightness, fluctuation level, skewness, and kurtosis at two different input EC powers: (a) $P_{EC}$ = 4 MW and (b) $P_{EC}$ = 2 MW. For skewness and kurtosis evaluations, the brightness data are first standardized such that each view's mean signal is zero and each standard deviation is one. The amplitude probability density function (PDF) is Gaussian with skewness of 0 and kurtosis of 0. First, in the high power case at $P_{EC}$ = 4 MW, the average brightness is maximum near the O-point region in columns 5-7, in line with the expected density and temperature profiles in the SOL (e.g., Figure 16 and 21 in [10]). The normalized fluctuation level is maximum near the last two outer columns, i.e., in the target shadow region. In both cases, the skewness level remains at a low level. In contrast, the far SOL region of the tokamaks generally exhibits a high level of skewness and kurtosis values, e.g., skewness ≲ 5 and kurtosis ≲ 40, reflecting highly intermittent, bursty turbulence behaviors [23]. Figure 8 shows the four probability distribution functions (PDFs) across the columns at the fixed 5th row (near the O-point) in the $P_{EC}$ = 4 MW case. A Gaussian fit curve, assuming a zero mean value, is overlaid as a blue dashed curve. In all cases, the experimental data closely match the Gaussian fit, showing no significant deviations except at high-amplitude events at $(I-\bar{I})/\sigma(I)$ > 4, where $\bar{I}$ is the time-averaged signal and $\sigma(I)$ is the standard deviation. This deviation begins approximately two decades below the maximum. These results are in line with the previous W7-X probe measurements [24].

One interesting observation is a skewness reversal in the main SOL region between the LCFS and the O-point at column 3, as shown in the third figure in Figure 8. Theory and modeling [25,26] discuss that this sign reversal of skewness is evidence of the curvature-driven instability. It is also theorized that the Kelvin-Helmholtz (KH) instability, an instability excited at the sheared flow boundary, plays a role [27]. ASDEX-Upgrade probe measurements report the birth of blobs and holes in the shear layer at the LCFS [28]. This skewness reversal is not universal in our observation, however, as it is absent at the third shear layer between the columns 6 and 7. Further, there is no reversal seen at the same second shear layer when the input power is reduced to 2 MW. It may be related to the weakening of the velocity gradient, while also indicating that the KH instability drive does not appear to be the dominant cause of all the shear layers.

Overall, the GPI data here serves as additional evidence that radial filamentary transport is not likely to be the dominant transport channel in the magnetic island region of the W7-X SOL, very different from what is found in the far-SOL region of tokamaks [5]. In future work, the analysis approach will be expanded to other magnetic configurations on W7-X to probe different parts of the SOL plasma. For example, the low iota configuration will allow probing the X-point region of the magnetic island, and the high mirror configuration will provide the GPI view to the far SOL region, offering broader insight into turbulence structures and dynamics in the 3D island SOL.

6. Summary

The study here provides an initial observation and characterization of turbulence characteristics and transport dynamics in the W7-X stellarator island divertor SOL plasma. High-spatial (~5 mm) and temporal (2 MHz) resolution GPI measurements in the O-point region of the magnetic island in the standard configuration are used to quantify the spatial structure of the flows and turbulence properties. The study shows that poloidal flows dominate over radial motion under the experimental conditions, with the poloidal velocity measured up to ~2 km/s and minimal radial motion (< 200 m/s), or nearly absent. Fluctuation statistics further support the lack of intermittent, bursty filamentary events, distinguishing the W7-X stellarator SOL behavior from those in tokamaks. The radial location of shear layers appears to align with regions of steep connection length transitions. In one case studied here, a skewness reversal is observed in the main SOL. These observations imply the 3D magnetic geometry influences flow dynamics and turbulence generation. Further studies will explore the underlying mechanisms and the interplay between the shear layers, connection length profile, and turbulence generation, as well as extend the investigation to other magnetic configurations on W7-X in order to gain a more comprehensive understanding of the turbulence transport process in the 3D island divertor.

7. Acknowledgment

Support for MIT participation was provided by the US Department of Energy, Fusion Energy Sciences, Award DE-SC0014251. This work has been carried out within the framework of the EUROfusion Consortium, funded by the European Union via the Euratom Research and Training Programme (Grant Agreement No 101052200 — EUROfusion). Views and opinions expressed are however those of the author(s) only and do not necessarily reflect those of the European Union or the European Commission. Neither the European Union nor the European Commission can be held responsible for them.